\documentstyle[12pt]{article}
\newcounter{saveeqn}
\title{
  {\vspace{-3cm} \normalsize \hfill
    \parbox{38mm}{MS-TPI-96-9 \\
                  cond-mat/9606091}  }\\[25mm]
  Calculation of Universal Amplitude Ratios\\
  in three-loop order
  }

\author{Christoph Gutsfeld\footnotemark[2],
        Jens K\"uster, Gernot M\"unster\\
        Institut f\"ur Theoretische Physik I,
        Universit\"at M\"unster\\
        Wilhelm-Klemm-Str.~9, D-48149 M\"unster, Germany\\
        e-mail: gutsfeld@physik.rwth-aachen.de,\\
        kuster@uni-muenster.de, munsteg@uni-muenster.de}
\date{June 18, 1996 (revised August 20, 1996)}
\begin{document}
\maketitle
\footnotetext[2]{Present address: Institut f\"ur
        Theoretische Physik E, RWTH Aachen, D-52056 Aachen}
\begin{abstract}

For the universality class of three-dimensional Ising systems the ratio of
the high- and low-temperature amplitudes for the correlation length and for
the susceptibility are universal quantities.  They can be calculated by
renormalized perturbation theory for scalar $\phi^4$ theory in fixed
dimensions $D=3$ in the symmetric phase and in the phase of broken symmetry. 
In this article the amplitude ratios are calculated in the three-loop
approximation.  Using the fixed point values of the coupling constants we
obtain $f_+/f_-=2.013(28)$ and $C_+/C_-=4.72(17)$.\\[5mm]
\end{abstract}

%
\section{Introduction}
Field-theoretic methods are a standard tool to obtain quantitative
results about the physics of second-order phase transitions.  In
particular they have been applied to the determination of universal
quantities like critical exponents or ratios of critical amplitudes.
These take the same values within large universality classes,
characterized by the dimensionality $D$ of space and the number $n$ of
components of the order parameter.

A quantity of theoretical and experimental interest is the ratio of
correlation length amplitudes $f_+ / f_-$.  It is defined by the
behaviour of the correlation length $\xi$ as a function of the
temperature $T$ near the critical temperature $T_c$ through
\begin{equation}
  \xi\sim\left\{\begin{array}{l@{,\quad}l}
      f_+t^{-\nu}&t>0\\f_-(-t)^{-\nu}&t<0\end{array}\right.;
  \quad t:=\frac{T-T_c}{T_c}\,.
\end{equation}
In this article we consider this amplitude ratio (as well as the
susceptibility amplitude ratio) for the universality class of the
three-dimensional Ising model.  The corresponding field-theoretical
model, which is assumed to be in the same class, is $\phi^4$ theory in
$D=3$ with a one-component ($n=1$) real scalar field.  The amplitude
ratio can be calculated by means of renormalized perturbation theory in
fixed dimensions $D=3$.  In a previous work \cite{MH94} the calculation
was done up to two loops.  For the theoretical background and more
details of the method we refer to that article and the references cited
therein.  In the present work we have extended the calculation to third
order of perturbation theory and introduced a new scheme to reduce the
relevant number of Feynman graphs.

The lagrangian density for the symmetric phase
($t>0$) is given by
\begin{eqnarray}
  {\cal L}(\phi_{0^+})&=&\frac{1}{2}\,(\partial\phi_{0^+}(x))^2
  +{\cal V}(\phi_{0^+})\nonumber\\
  {\cal V}(\phi_{0^+})&=&\frac{1}{2}\,m_{0^+}^2\phi^2_{0^+}(x)
  +\frac{1}{4!}\,g_0\phi^4_{0^+}(x).
\end{eqnarray}

For the phase of broken symmetry ($t<0$) we use a shift in the
field variable and expand the potential around the minimum value
$v_0:=\sqrt{3m^2_{0^-}/g_0}$ omitting constant terms.
\begin{equation}
  \phi_{0^-}(x):=\phi(x)-v_0,
\end{equation}
\begin{eqnarray}
  {\cal L}(\phi_{0^-})&=&\frac{1}{2}\,(\partial\phi_{0^-}(x))^2
  +{\cal V}(\phi_{0^-})\nonumber\\
  {\cal V}(\phi_{0^-})&=&\frac{1}{2}\,m^2_{0^-} \phi^2_{0^-}(x)
  +\frac{1}{3!}\sqrt{3g_0}\,m_{0^-}\phi^3_{0^-}(x)
  +\frac{1}{4!}\,g_0\phi^4_{0^-}(x)\,.
\end{eqnarray}
Besides the $\phi^4$ self-interaction there is a cubic $\phi^3$
interaction.  This term gives rise to a large number of additional
Feynman graphs, many of them containing tadpole subgraphs.  To simplify
the calculation and to reduce the number of graphs, we have used a
method to eliminate tadpole subgraphs by means of a modification of the
mass and coupling parameters \cite{KM96}.

On the three-loop level no new divergencies arise.  So we only have to
handle subdivergencies, which are isolated by dimensional regularization.

In the next section we summarize the renormalization scheme.  The
results of the perturbative calculations are presented in section 3, and
the numerical estimates are discussed in section 4.

%
\section{Renormalized perturbation theory}
The correlation length considered here is defined through the second
moment of the connected two-point function.
\begin{equation}
  \xi^2:=\frac{1}{2D}\,\frac{\int d^Dx\,x^2 G^{(2,0)}_c(x)}
  {\int d^Dx\,G^{(2,0)}_c(x)}
  =-\frac{\frac{\partial}{\partial p^2}\,G^{(2,0)}_c(p)}
  {G^{(2,0)}_c(p)}\,\Bigg|_{p=0} \,,
\end{equation}
\begin{equation}
  G^{(2,0)}_c(x):=\langle\phi_0(x)\phi_0(0)\rangle
  -\langle\phi_0(x)\rangle\langle\phi_0(0)\rangle\,.
\end{equation}
The two-point function is related to the two-point vertex function
$\Gamma^{(2,0)}$ by
\begin{equation}
  -\Gamma_0^{(2,0)}(p)=\left(G^{(2,0)}_c(p)\right)^{-1} \,.
\end{equation}
The renormalized mass $m_R$ is defined by
\begin{equation}
  m^2_R:=\frac{\Gamma_0^{(2,0)}(p)}{\frac{\partial}{\partial p^2}\,
    \Gamma_0^{(2,0)}(p)}\,\Bigg|_{p=0}
  =\frac{1}{\xi^2}\,,
\end{equation}
and coincides with the inverse of the correlation length.

As $-\Gamma_0^{(2,0)}=m_0^2+p^2-\Sigma$, where $\Sigma(p)$
is the sum of all one-particle irreducible two-point graphs
with amputated external legs, we can determine the correlation length
$\xi$ perturbatively.

The renormalization constants are defined through
\begin{equation}
  Z_3^{-1}:= -\frac{\partial\Gamma_0^{(2,0)}(p;m_0,g_0)}
  {\partial p^2}\,\bigg|_{p=0}
\end{equation}
and
\begin{equation}
  Z_2^{-1}:= -\Gamma_0^{(2,1)}(\{0;0\};m_0,g_0)
  = -\frac{\partial}{\partial m_0^2}\,\Gamma_0^{(2,0)}
  (0;m_0,g_0)\,.
\end{equation}

In order to obtain the universal amplitude ratio $f_+ / f_-$ various
quantities have to be expanded in powers of a dimensionless renormalized
coupling $u_R$. The following renormalization schemes are used in the
two different phases.
\subsection{Symmetric phase}
The renormalized coupling constant in the symmetric phase $g_R^{(4)}$
is, following \cite{A84,MM94}, defined by the value of the
four-point vertex function at zero external momenta.
\begin{equation}
  g_R^{(4)}:= -Z_3^2\Gamma_0^{(4,0)}(\{0\};m_0,g_0)\,.
\end{equation}
It is related to the dimensionless renormalized coupling $u_R$ by
\begin{equation}
  g_R^{(4)} = m_R^{4-D}u_R\,.
\end{equation}
Additionally we define the coupling renormalization constant
\begin{equation}
  Z_1^{-1}:= -\frac{1}{g_0}\,\Gamma_0^{(4,0)}(\{0\};m_0,g_0) \,.
\end{equation}

The complete set of renormalization conditions are
\stepcounter{equation}\setcounter{saveeqn}
{\value{equation}}
\setcounter{equation}{0}
\renewcommand{\theequation}
{\mbox{\arabic{saveeqn}\alph{equation}}}
\begin{eqnarray}
  \label{rensym}
  \Gamma_R^{(2,0)}(0;m_R,u_R)&=&-m_R^2 \\
  \frac{\partial}{\partial p^2}\,\Gamma_R^{(2,0)}(p;m_R,u_R)\,
  \bigg|_{p^2=0}&=&-1 \\
  \Gamma_R^{(4,0)}(\{0\};m_R,u_R)&=&-g_R^{(4)} \\
  \Gamma_R^{(2,1)}(\{0;0\};m_R,u_R)&=&-1\,.
\end{eqnarray}
\setcounter{equation}{\value{saveeqn}}
\renewcommand{\theequation}{\arabic{equation}}
\subsection{Phase of broken symmetry}
In the phase of broken symmetry the field has a non-vanishing
expectation value
\begin{equation}
  v=v_0+G^{(1,0)}_c,\quad \mbox{where} \quad v_0=\sqrt{3m_0^2/g_0}\,.
\end{equation}
The renormalized field expectation value is
\begin{equation}
  v_R:=\frac{1}{\sqrt{Z_3}}\,v \,.
\end{equation}
We define the renormalized coupling constant $g_R$ through the one-point
function according to \cite{LW87}
\begin{equation}
  g_R:=\frac{3m_R^2}{v_R^2}\,.
\end{equation}
The dimensionless coupling $u_R$ is again introduced by
\begin{equation}
  g_R = m_R^{4-D}u_R\,.
\end{equation}
The renormalization scheme in this phase is thus summarized by
\stepcounter{equation}\setcounter{saveeqn}
{\value{equation}}
\setcounter{equation}{0}
\renewcommand{\theequation}
{\mbox{\arabic{saveeqn}\alph{equation}}}
\begin{eqnarray}
  \label{renbro}
  \Gamma_R^{(2,0)}(0;m_R,u_R)&=&-m_R^2 \\
  \frac{\partial}{\partial p^2}\,\Gamma_R^{(2,0)}(p;m_R,u_R)\,
  \bigg|_{p^2=0}&=&-1 \\
  \frac{3 m_R^2}{v_R^2}&=&m_R^{4-D}u_R\, =\,g_R \\
  \Gamma_R^{(2,1)}(\{0;0\};m_R,u_R)&=&-1\,.
\end{eqnarray}
\setcounter{equation}{\value{saveeqn}}
\renewcommand{\theequation}{\arabic{equation}}

%
\section{Perturbation series}
As the method was already developed in \cite{MH94} the main problem in
the present calculation was the number of diagrams in third order of
perturbation theory.  According to the renormalization conditions
(\ref{rensym}-d, \ref{renbro}-d) we had to calculate the two-point
vertex function and its momentum derivative at zero momentum in both
phases.  In the symmetric phase the four-point function at zero momenta
and in the phase of broken symmetry the vacuum expectation value of the
field had to be calculated, too.  {}From these series the masses and
coupling constants in both phases are derived.  In order to distinguish
the parameters in the two phases we label them with an index $+$ for the
symmetric (high temperature) and $-$ for the broken-symmetric (low
temperature) phase.  The natural expansion variables are the
dimensionless renormalized couplings
\begin{equation}
  u_{R+}:=\frac{g_{R+}^{(4)}}{m_{R+}^{4-D}} \makebox[2cm]{and}
  u_{R-}:=\frac{g_{R-}}{m_{R-}^{4-D}}\,.
\end{equation}
The bare dimensionless coupling is defined analogously with the bare
parameters
\begin{equation}
  u_0:=\frac{g_0}{m_0^{4-D}} \,.
\end{equation}

The number of diagrams we encountered is already non-negligible: there
are 204 one-particle irreducible diagrams contributing to the inverse
propagator at the three-loop level, compared to 20 at two loops.
After exploiting symmetries there are still 162 of them.
Many of them contain tadpoles and their elimination by means of
Dyson-Schwinger equations simplifies the book-keeping very much
\cite{KM96}. The reduced set of one-particle irreducible
propagator-diagrams without tadpoles only contains 34 elements.
Nevertheless we checked the calculation by means of the usual
perturbation theory.
The program QGRAF by P.\ Nogueira \cite{NOG93} was helpful in verifying the
completeness of our list of diagrams.

The starting point of the calculation are the expansions of the
renormalized masses and couplings in terms of the bare coupling.
For this purpose a regularization scheme has to be used. The final
results are independent of the choice of the regularization scheme.
We decided to use dimensional regularization with $D=3-\epsilon$.

In the symmetric phase the expansions are
\begin{eqnarray}
  m_{R+}^2 & = & m_0^2 \left[1-
    \frac{u_0}{8\pi}\left(1-{\textstyle\frac{\epsilon}{2}}
      \left(\gamma+\ln{\textstyle\frac{m_0^2}{\pi}}-2\right)
      +{\cal O}(\epsilon^2)\right)
      + \left(\frac{79}{162}-\frac{1}{3}B^{div} \right)
      \left(\frac{u_0}{8\pi}\right)^2 \right. \nonumber \\
  & & \left.+\left(-\frac{131}{216}+\frac{71}{27}\ln\frac{4}{3}
      +\frac{32}{3}a+\frac{1}{6}B_1^{div}\right)
    \left(\frac{u_0}{8\pi}\right)^3+{\cal O}\left(u_0^4\right)\right],\\
  g_{R+}^{(4)} & = & g_0\left[1-\frac{3}{2}\frac{u_0}{8\pi}
    \left(1-{\textstyle\frac{\epsilon}{2}}\left(\gamma
      +\ln{\textstyle\frac{m_0^2}{\pi}}\right)
      +{\cal O}(\epsilon^2)\right)
    +\left(-\frac{2}{81}+2(1+{\cal O}(\epsilon))\right)
    \left(\frac{u_0}{8\pi}\right)^2\right. \nonumber \\
  & & \left.+\left(\frac{199}{1296}-\frac{373}{54}\ln\frac{4}{3}
      -\frac{128}{3}a-C^{\mbox{\scriptsize\it Tet}}
      -\frac{1}{4}B_1^{div}\right)
    \left(\frac{u_0}{8\pi}\right)^3
    +{\cal O}\left(u_0^4\right)\right], \\
  u_{R+} & = & u_0 \left[1-\frac{u_0}{8\pi}
    \left(1-{\textstyle\frac{\epsilon}{2}}
      \left(\gamma+\ln{\textstyle\frac{m_0^2}{\pi}}+2\right)
      +{\cal O}(\epsilon^2)\right)
      +\left(\frac{329}{216}+\frac{1}{6}B^{div}\right)
      \left(\frac{u_0}{8\pi}\right)^2\right. \nonumber \\
  & & \left.+\left(\frac{13}{9}-\frac{74}{9}\ln\frac{4}{3}-48a
      -C^{\mbox{\scriptsize\it Tet}}-\frac{1}{3}B_1^{div}\right)
      \left(\frac{u_0}{8\pi}\right)^3
      +{\cal O}\left(u_0^4\right)\right]\,.
\end{eqnarray}
To first order in the coupling constant $u_0$ we have to keep terms of
order $\epsilon$, because in second order there are divergent terms
$B^{div}$ which are proportional to $\epsilon^{-1}$. By multiplication
they give finite contributions in the third order.

The divergent terms $B^{div}$ and $B_1^{div}$ cancel out in the final
results and need not be displayed here.
$\gamma$ is Euler's constant, and the other constants used in these
equations are:
\begin{equation}
  a = \frac{\pi^2}{48} - \frac{1}{8}\ln^2\left({\textstyle\frac{4}{3}}
  \right) - \frac{1}{3}\ln\left({\textstyle\frac{4}{3}}\right)
  - \frac{1}{4}\mbox{Li}_2\left({\textstyle\frac{1}{4}}\right)
  = 0.0324645
\end{equation}
\cite{BS92} with the dilogarithm
\begin{equation}
  \mbox{Li}_2(x) = - \int_0^x \!dt\,\frac{\ln (1-t)}{t} \,,
\end{equation}
and
\begin{eqnarray}
  C^{\mbox{\scriptsize\it Tet}} & = & \pi^{-6} \int d^3k_1 d^3k_2 d^3k_3
  \Delta(k_1) \Delta(k_2) \Delta(k_3) \Delta(k_1-k_2) \Delta(k_2-k_3)
  \Delta(k_3-k_1) \nonumber \\
  & = & 0.1739006\,,
\end{eqnarray}
where $\Delta(k) = (k^2+1)^{-1}$.
This is the only integral which we could not solve analytically.
The numerical value stems from \cite[graph 12U4]{BNM77} and was
confirmed by our own calculations.

In the broken-symmetry phase we get
\begin{eqnarray}
  m_{R-}^2 & = & m_0^2 \left[1+ \frac{3}{8}\frac{u_0}{8\pi}
    \left(1-{\textstyle\frac{\epsilon}{2}}\left(\gamma
        +\ln{\textstyle\frac{m_0^2}{\pi}}-10\right)
      +{\cal O}(\epsilon^2)\right)
      + \left(\frac{3973}{5184}+\frac2 3 B^{div} \right)
      \left(\frac{u_0}{8\pi}\right)^2 \right. \nonumber \\
  & & \left.+\left(-\frac{101245}{41472}+\frac{21535}{2592}
      \ln\frac{4}{3}-\frac{1723}{48}a
      -\frac{3345}{1024} C^{\mbox{\scriptsize\it Tet}}
      +\frac{1}{8}B_1^{div}\right)
      \left(\frac{u_0}{8\pi}\right)^3
    +{\cal O}\left(u_0^4\right)\right], \nonumber\\ \\
  g_{R-} & = &
  g_0\left[1-\frac{7}{4}\frac{u_0}{8\pi}\left(1
      -{\textstyle\frac{\epsilon}{2}}\left(\gamma
        +\ln{\textstyle\frac{m_0^2}{\pi}}
        {\textstyle -\frac{25}{56}}\right)
          +{\cal O}(\epsilon^2)\right)+\frac{17099}{5184}
    \left(\frac{u_0}{8\pi}\right)^2\right. \nonumber \\
  & & \left.+\left(-\frac{4051}{576}
      +\frac{21319}{1296}\ln\frac{4}{3}-\frac{1045}{24}a
      -\frac{2849}{512}C^{\mbox{\scriptsize\it Tet}}
      +\frac{7}{12}B_1^{div}\right)
    \left(\frac{u_0}{8\pi}\right)^3
    +{\cal O}\left(u_0^4\right)\right], \nonumber\\ \\
  u_{R-} & = &
  u_0 \left[1-\frac{31}{16}\frac{u_0}{8\pi}\left(1
      -{\textstyle\frac{\epsilon}{2}}\left(\gamma
        +\ln{\textstyle\frac{m_0^2}{\pi}}{\textstyle
          -\frac{44}{31}}\right)+{\cal O}(\epsilon^2)\right)
    +\left(\frac{40957}{13824}-\frac{1}{3}B^{div}\right)
    \left(\frac{u_0}{8\pi}\right)^2\right.\nonumber \\
  & & \left. +\left(-\frac{284719}{73728}+\frac{21247}{1728}
      \ln\frac{4}{3}-\frac{819}{32}a
      -\frac{8051}{2048}C^{\mbox{\scriptsize\it Tet}}
      +\frac{31}{24}B_1^{div}\right)\left(\frac{u_0}{8\pi}\right)^3
    +{\cal O}\left(u_0^4\right)\right].\nonumber \\
\end{eqnarray}

For the calculation of the universal amplitude ratio of correlation
lengths we need the functions
\begin{equation}
  F_{\pm}(u_{R^{\pm}}):=\frac{\partial m^2_{R^{\pm}}}
  {\partial m^2_{0^{\pm}}}\,\bigg|_{g_0}.
\end{equation}
When the bare masses and couplings are expressed in terms of the
renormalized ones the perturbation series are
\begin{eqnarray}
  F_+(u_{R+}) & = &
  1-\frac{1}{2}\frac{u_{R+}}{8\pi}
  -\frac{1}{6}\left(\frac{u_{R+}}{8\pi}\right)^2 \nonumber \\
  & & +\left(\frac{13}{27}-\frac{71}{54}\ln\frac{4}{3}
    -\frac{16}{3}a\right) \left(\frac{u_{R+}}{8\pi}\right)^3
  +{\cal O}\left(u_{R+}^4\right) \,, \\
  F_-(u_{R-}) & = & 1+\frac{3}{16}\frac{u_{R-}}{8\pi}
  -\frac{233}{768}\left(\frac{u_{R-}}{8\pi}\right)^2 \nonumber \\
  & & +\left(-\frac{297256}{663552}-\frac{21535}{5184}
    \ln\frac{4}{3}+\frac{1723}{96}a
    +\frac{3345}{2048}C^{\mbox{\scriptsize\it Tet}}\right)
  \left(\frac{u_{R-}}{8\pi}\right)^3
  +{\cal O}\left(u_{R-}^4\right)\,. \nonumber \\
\end{eqnarray}

At this point we still have two coupling constants, $u_{R+}$ and
$u_{R-}$.  Following \cite{MH94} we introduce a new coupling $\bar
u_R$, which is defined in both phases, such that the corresponding
$\beta$-functions in both phases and consequently the numerical values
of the fixed point couplings are equal.  Previous experience suggests to
choose $\bar u_R$ such that it coincides with the usual coupling in the
low-temperature phase:  $\bar u_R \equiv u_{R-}$.  Therefore we
generally denote it by $u_{R-}$ in the following.  The other coupling
constant expressed as a series in $u_{R-}$ is
\begin{eqnarray}
  u_{R+}(u_{R-}) & = &
  u_{R-}\left[1+\frac{1}{4}\frac{u_{R-}}{8\pi}+\frac{1633}{2592}
    \left(\frac{u_{R-}}{8\pi}\right)^2\right. \nonumber \\
  & & \left.+\left(\frac{1011239}{165888}-\frac{30271}{1296}
      \ln\frac{4}{3} +\frac{7}{8}a
      +\frac{2337}{512}C^{\mbox{\scriptsize\it Tet}}\right)
    \left(\frac{u_{R-}}{8\pi}\right)^3
    +{\cal O}\left(u_{R-}^4\right)\right] \,. \nonumber \\
\end{eqnarray}
This relation allows to express all renormalized perturbation series in
terms of a single coupling constant $u_{R-}$.
In particular for the ratio $F_{-} / F_{+}$ we obtain
\begin{eqnarray}
  \Phi_-(u_{R-}) & := &
  \frac{F_-\left(u_{R^-}\right)}{F_+\left(u_{R^+}\right)} \nonumber \\
  & = & 1+\frac{11}{16}\frac{u_{R-}}{8\pi}+\frac{85}{256}
  \left(\frac{u_{R-}}{8\pi}\right)^2 \nonumber \\
  & & +\left(-\frac{109217}{663552}-\frac{14719}{5184}\ln\frac{4}{3}
    +\frac{745}{32}a
    +\frac{3345}{2048}C^{\mbox{\scriptsize\it Tet}}\right)
  \left(\frac{u_{R-}}{8\pi}\right)^3
  +{\cal O}\left(u_{R-}^4\right). \nonumber \\
\end{eqnarray}
This function finally yields the desired amplitude ratio via
\begin{equation}
  \label{fvonFnu}
  \frac{f_+}{f_-} =
  \left[2 \Phi_-(u_{R-}^*) \right]^{\nu} \,,
\end{equation}
where $u_{R-}^*$ is the fixed point value of the coupling and $\nu$ is
the correlation length exponent.
Both $u_{R-}^*$ and $\nu$ can be obtained in perturbation theory too,
but more precise values are available in the literature and we shall
make use of them.

%
\section{Numerical results}
What is needed for the amplitude ratio of correlation lengths is the
value of $\Phi_-(u_{R-})$ at $u_{R-} = u_{R-}^*$.  For the fixed
point $u_{R-}^*$ we take an estimate $u_{R-}^* = 14.73(14)$ from
low-temperature series \cite{S93} and another estimate $u_{R-}^* =
15.1(1.3)$ used in \cite{M90}.  For comparison we note that the zero of
the 3-loop $\beta$-function is located at $u_{R-}^* = 14.2$, using a
(2,1)-Pad\'{e}-Borel approximation.
\begin{table}[htbp]
  \begin{center}
    \leavevmode
    \begin{tabular}{c|cccc}
      \hline Fixed point & \multicolumn{4}{c}{$\Phi_-(u_{R-}^*)$}\\
      \cline{2-5}  $u_{R-}^*$ &
      [3,0]-Pad\'e & [2,1]-Pad\'e & [1,2]-Pad\'e & [0,3]-Pad\'e \\ \hline
      14.73 \cite{S93} & 1.5288 & 1.5301 & 1.5002 & 1.5149 \\ \hline
      15.1  \cite{M90} & 1.5456 & 1.5471 & 1.5133 & 1.5301 \\ \hline
    \end{tabular}
    \caption{$\Phi_-$ as a function of the low-temperature fixed point
      $u_{R-}^*$}
  \end{center}
\end{table}
The entries in table 1 result from the possible Pad\'e approximants to
$\Phi_-(u_{R-})$ evaluated at $u_{R-}^*$. The values are quite close
together and the mean value is 1.526(26), where the error represents the
maximal deviation. The numerical convergence of the series at the
fixed point is rather good,
\begin{equation}
\Phi_-(u_{R-}^*) = 1 + 0.410 + 0.118 + 0.012\,,
\end{equation}
although it is expected to be asymptotic only. An application of the
usual Pad\'e-Borel summation method does not improve the result. On the
contrary the uncertainty is more than doubled.

Using the high-temperature coupling ($u_{R+}^* \approx 24$) as an
expansion parameter is much worse.  For $\Phi_+(u_{R+})$ we get a mean
value of 1.39(34) with Pad\'e, and 1.48(24) with Pad\'e-Borel summation,
respectively, so the maximal deviation is more than ten times higher as
above. This is related to the poor numerical convergence:
\begin{equation}
\Phi_+(u_{R+}^*) = 1 + 0.657 + 0.146 - 0.396\,.
\end{equation}
Therefore we consider the estimate from $\Phi_-$ as more reliable.

For the critical exponent $\nu$ we used Monte Carlo results
($\nu=0.624(2)$) \cite{BGHP92} and values of renormalized perturbation
theory ($\nu=0.6300$) \cite{LZ80}.  After exponentiation with these
values for $\nu$ we get the universal amplitude ratio for the
correlation length $\xi$ according to (\ref{fvonFnu}),
\begin{equation}
  \label{Qfend}
  \frac{f_+}{f_-} = 2.013(28) \,.
\end{equation}

Using the high-temperature coupling instead we would get a value of
1.98(20) employing Pad\'e-Borel approximations.  The error is ten times
larger than in (\ref{Qfend}).  As discussed above the low-temperature
coupling appears to be the better expansion parameter.

With a similar method we have calculated the universal amplitude ratio
of the susceptibility.  As before we get best results with the
low-temperature coupling, namely
\begin{equation}
  \frac{C_+}{C_-} = 4.72(17) \,.
\end{equation}
The same ratio
has been calculated by means of three-dimensional perturbation theory in
\cite{BB87} with the result $C_+/C_-=4.77(30)$.

%
\section{Conclusion}
Our third order calculation of the amplitude ratio of correlation
lengths (\ref{Qfend})
is a confirmation and improvement of the two-loop result of 2.03(4)
\cite{MH94}.  Theoretical estimates in the $\epsilon$-expansion
(1.91 \cite{BLZ74}), high- and low-temperature expansions
(1.96(1) \cite{LF89}, 1.94(3) \cite{S93}) are lower. Experimental values
(2.05(22), 2.22(5) \cite{KKG83}, 1.9(2), 2.0(4) \cite{PHA91}) and Monte
Carlo results (2.06(1) \cite{RWZ94}) are above or close to our results.

\vspace{6mm}
\noindent
{\large\bf Acknowledgement}: We thank D.\ Broadhurst and P.\ Nogueira for
discussions.

%
\end{document}